\documentclass[twocolumn,showpacs,superscriptaddress,preprintnumbers,amsmath,amssymb,aps,prb]{revtex4-1}
\usepackage{graphicx}
\usepackage{mathptmx}
\usepackage{verbatim}
\usepackage{graphicx}
\usepackage{amsmath}
\usepackage{amssymb}
\usepackage{amsfonts}
\usepackage{mathrsfs}
\usepackage{amsmath}     %  needed for subequations
\usepackage{color}
\usepackage{graphicx}
\usepackage{bm}               %  bold  maths  \Delta_0
\usepackage{amssymb}
\usepackage{xspace} 
\usepackage{hyperref}
\usepackage{multirow}
\usepackage{color}
\usepackage{float}
\usepackage{bbold}
\usepackage{tabu}
\usepackage{textcomp}
\usepackage{multirow}
\usepackage{dcolumn}     %    Align table columns on decimal point
\usepackage{appendix} %%     

\newcolumntype{L}[1]{>{\raggedright\arraybackslash}m{#1}}
\newcolumntype{C}[1]{>{\centering\arraybackslash}m{#1}}
\newcolumntype{R}[1]{>{\raggedleft\arraybackslash}m{#1}}
\newcolumntype{N}{@{}m{0pt}@{}}

\usepackage{listings}
\usepackage{color}

\definecolor{dkgreen}{rgb}{0,0.6,0}
\definecolor{gray}{rgb}{0.5,0.5,0.5}
\definecolor{mauve}{rgb}{0.58,0,0.82}

\begin{document}

\title{Open-orbit induced low field extremely large magnetoresistance in graphene/h-BN superlattices}

\author{Zihao Wang}
\email{zihao.wang@nus.edu.sg}
\affiliation{Institute for Functional Intelligent Materials, NUS, Block S9, Level 9, 4 Science Drive, Singapore 117544, Singapore}
\affiliation{Department of Materials Science and Engineering, National University of Singapore, 117544, Singapore}

\author{Pablo M. Perez-Piskunow}
\email{pablo.perez.piskunow@gmail.com}
\affiliation{Catalan Institute of Nanoscience and Nanotechnology - ICN2, (CSIC and BIST), Campus UAB, Bellaterra, Barcelona 08193, Spain}

\author{Calvin Pei Yu Wong}
\affiliation{Institute of Materials Research and Engineering (IMRE), Agency for Science Technology and Research (A*STAR), 2 Fusionopolis Way, Innovis \#08-03, Singapore 128634, Republic of Singapore}

\author{Matthew Holwill}
\affiliation{Department of Physics and Astronomy, University of Manchester, Oxford Road, Manchester M13 9PL, UK}

\author{Jiawei Liu}
\affiliation{Institute for Functional Intelligent Materials, NUS, Block S9, Level 9, 4 Science Drive, Singapore 117544, Singapore}

\author{Wei Fu}
\affiliation{Institute of Materials Research and Engineering (IMRE), Agency for Science Technology and Research (A*STAR), 2 Fusionopolis Way, Innovis \#08-03, Singapore 128634, Republic of Singapore}

\author{Junxiong Hu}
\affiliation{Department of Physics, National University of Singapore, 2 Science Drive 3, Singapore 117551, Republic of Singapore}

\author{T Taniguchi}
\affiliation{National Institute for Materials Science, Ibaraki 305-0044, Japan}

\author{K Watanabe}
\affiliation{National Institute for Materials Science, Ibaraki 305-0044, Japan}

\author{Ariando Ariando}
\affiliation{Department of Physics, National University of Singapore, 2 Science Drive 3, Singapore 117551, Republic of Singapore}

\author{Lin Li}
\affiliation{Institute for Functional Intelligent Materials, NUS, Block S9, Level 9, 4 Science Drive, Singapore 117544, Singapore}

\author{Kuan Eng Johnson Goh}
\affiliation{Institute of Materials Research and Engineering (IMRE), Agency for Science Technology and Research (A*STAR), 2 Fusionopolis Way, Innovis \#08-03, Singapore 128634, Republic of Singapore}
\affiliation{Department of Physics, National University of Singapore, 2 Science Drive 3, Singapore 117551, Republic of Singapore}

\author{Stephan Roche}
\email{stephan.roche@icn2.cat}
\affiliation{Catalan Institute of Nanoscience and Nanotechnology - ICN2, (CSIC and BIST), Campus UAB, Bellaterra, Barcelona 08193, Spain}
\affiliation{ICREA Institucio Catalana de Recerca i Estudis Avancats, Barcelona 08010, Spain}

\author{Jeil Jung}
\email{jeiljung@uos.ac.kr}
\affiliation{Department of Physics, University of Seoul, Seoul 02504, Korea}
\affiliation{Department of Smart Cities, University of Seoul, Seoul 02504, Korea}

\author{Konstantin Novoselov}
\email{kostya@nus.edu.sg}
\affiliation{Institute for Functional Intelligent Materials, NUS, Block S9, Level 9, 4 Science Drive, Singapore 117544, Singapore}
\affiliation{Department of Materials Science and Engineering, National University of Singapore, 117544, Singapore}

\author{Nicolas Leconte}
\email{lecontenicolas0@uos.ac.kr}
\affiliation{Department of Physics, University of Seoul, Seoul 02504, Korea}

\date{\today}
\begin{abstract}
We report intriguing and hitherto overlooked low-field room temperature extremely large magnetoresistance (XMR) patterns in graphene/hexagonal boron nitride (h-BN) superlattices that emerge due to the existence of open orbits within each miniband. This finding is set against the backdrop of the experimental discovery of the Hofstadter butterfly in moiré superlattices, which has sparked considerable interest in the fractal quantum Hall regime. To cope with the challenge of deciphering the low magnetic field dynamics of moiré minibands, we utilize a novel semi-classical calculation method, grounded in zero-field Fermi contours, to predict the nontrivial behavior of the Landau-level spectrum. This is compared with fully quantum simulations, enabling an in-depth and contrasted analysis of transport measurements in high-quality graphene-hBN superlattices. Our results not only highlight the primary observation of the open-orbit induced XMR in this system but also shed new light on other intricate phenomena. These include the nuances of single miniband dynamics, evident through Lifshitz transitions, and the complex interplay of semiclassical and quantum effects between these minibands. Specifically, we document transport anomalies linked to trigonal warping, a semiclassical deviation from the expected linear characteristics of Landau levels, and magnetic breakdown phenomena indicative of quantum tunneling, all effects jointly contributing to the intricacies of a rich electronic landscape uncovered at low magnetic fields.
\end{abstract}
\maketitle

\section*{Introduction}

The revelation of a fractal electronic spectrum in graphene superimposed on hexagonal boron nitride (hBN)~\cite{Dean2013,Ponomarenko2013,Hunt:2013ef}  marked the inception of a decade-long research odyssey into the properties of the Hofstadter butterfly~\cite{PhysRevB.14.2239}. This research endeavor has primarily revolved around the strategic introduction of superlattices in pristine graphene, leading to the intricate division of its intrinsic $\pi-$electron bands into a series of moirÃ© minibands~\cite{Dean2013,Ponomarenko2013,Hunt:2013ef}. The fractal's self-similar nature, as posited in the original Hofstadter model~\cite{PhysRevB.14.2239}, suggests an endless emergence of spectrum replica features. However, this pursuit is constrained physically by some fundamental mechanisms that are disregarded in the idealized mathematical fractal model. These mechanisms, either
resulting from the dynamics within individual minibands or from interactions between
different minibands, present distinct experimental phenomena acting on high-energy minibands at low magnetic fields.
Over the past decade, research on graphene's Landau levels has predominantly concentrated on high-field and low-energy regions. Indeed, because the onset of the quantum hall effect is influenced by disorder, high-energy minibands have been challenging to map clearly due to high requirements in measurement precision and due to the too large samples that are needed for theoretical simulation at low fields.~\cite{woods, Amet:2013gw,Hunt:2013ef,Wang:2016ii,Yankowitz:2014bv,doi:10.1126/science.1237240,0a7c5baa122e4f1f9f96b95b0f35eb3e,Wallbank:2013ep,Chen2014,Ni2019,PhysRevLett.111.266801,RibeiroPalau2018,PhysRevB.90.075428,10.1021/acs.nanolett.8b03423,PhysRevB.100.195413,Long2022}. 

Here, by fabricating ultraclean graphene superlattice samples that minimize disorder and conducting ultra-high-precision transport measurements that mitigate thermal noise, we have obtained a clear experimental mapping in the low-field and high-energy regions. Meanwhile, theoretically using an innovative semi-classical calculation technique - which exploits zero-field calculations to predict behavior under finite magnetic fields, hence bypassing the impracticality of too large samples required for full quantum simulations - we embark on a systematic and comprehensive analysis of the electronic behavior in the same regime. We unravel a knowledge frontier that goes beyond the limits of fractality, entering a region where Lifshitz transitions play a pivotal role. Specifically, the experimental manifestation of open orbits at saddle point vHS is seminally identified as XMR persisting up to room temperature, and the Landau level splitting between the hole-side Secondary Dirac Point (h-SDP) and the neighboring vHS is attributed to trigonal warping captured by the lattice-reconstructed tight-binding model. We finally also identify a semi-classical deviation from the linear characteristics typically expected for Landau levels, as evidenced in the Wannier representation of the Hofstadter diagram, and observe signatures of magnetic breakdown, indicative of quantum tunneling phenomena.

\section*{G-hBN: full spectrum}

\begin{figure*}[tbhp]%
\centering
\includegraphics[width=1.0\textwidth]{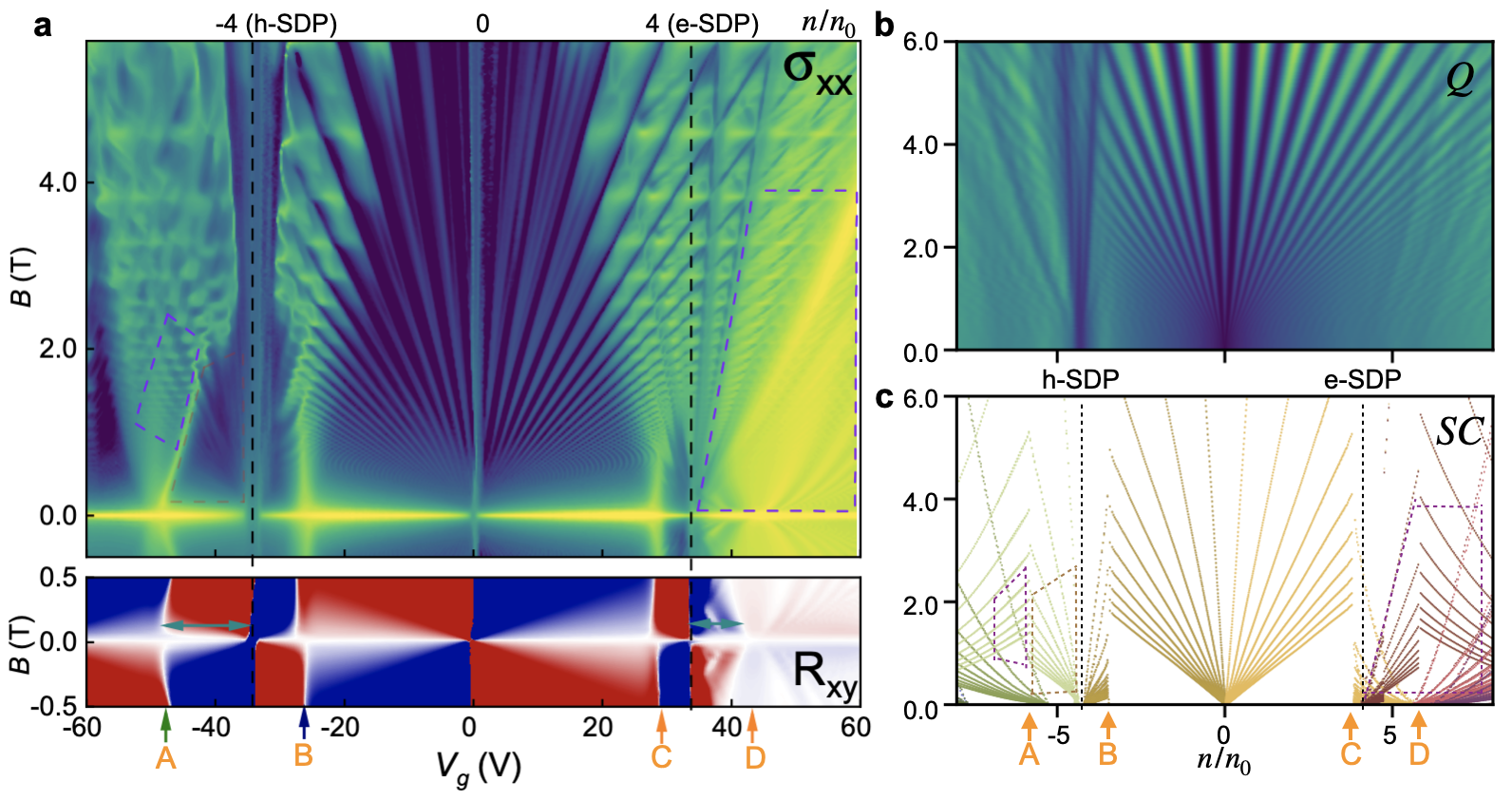}
\caption{Full spectrum magnetic-field dependence of electronic states in G-hBN. \textbf{a,} $\sigma_\text{xx}$ (top) and $R_\text{xy}$ (bottom) transport measurements. The black dashed line represents the position of SDP for hole  and electron side,respectively. The region enclosed by grey dashed lines labels the region where Landau level splitting occurs. The region enclosed by the purple dashed lines points to the place where the non-linear Wannier diagram appears. The four vHS positions, manifested as conductivity dips and Rxy sign reversal, are denoted by \textbf{A}, \textbf{B}, \textbf{C}, and \textbf{D}. \textbf{b,}. Fully quantum ($Q$) numerical charge-carrier-dependent capacitance calculation using Lanczos recursion. \textbf{c,} Charge-carrier dependent semi-classical ($SC$) LL predictions using the zero-field Fermi contour behavior. Each color represents the LLs belonging to a specific minibands. 
}
\label{fig2}
\end{figure*}

In Fig.~\ref{fig2}\textbf{a}, we present the full spectrum of the high-resolution transport $\sigma_\text{xx}$ and $R_\text{xy}$ measurements on the graphene-hBN superlattice. In addition to the usual Hofstadter butterfly signatures, we observe a variety of new features in the $[0-5]$T regime at the onset of the Quantum Hall Effect that have been overlooked in past experiments. In the $R_\text{xy}$ map, the positions of the vHS denoted as \textbf{A}, \textbf{B}, \textbf{C}, and \textbf{D} are determined by a sign reversal, corresponding to the conductivity dips in the $\sigma_\text{xx}$ map. In the $R_\text{xy}$ map, the blue double arrows mark the distance between the vHS and SDP on both sides, indicating the asymmetry between the electron and hole side that leads to markedly different observations. Both to the left of vHS \textbf{A} and around vHS \textbf{D}, a surprising reshaping of the linear Landau levels is observed in the regions enclosed by purple dashed lines. Indeed, a non-linearity occurs along with their broadening as the magnetic field increases. Additionally, between the vHS \textbf{A} and the h-SDP in the region enclosed by dashed grey lines, the linear Landau levels split. All these phenomena revolve around vHS, a type of Lifshitz transition, or within regions where bands overlap, and thus cannot be explained by the single isolated band theory used for the Hofstadter butterfly.
In the right panels, both the quantum (Fig.~\ref{fig2}\textbf{b}) and the semi-classical (Fig.~\ref{fig2}\textbf{c}) charge carrier density-dependent pictures are showcased through the integration of the energy-dependent DOS from Fig.~\ref{fig1}\textbf{d}. The semi-classical simulation agrees well with the non-linear Landau levels observed in Fig.~\ref{fig2}\textbf{a}. Furthermore, the crossing of Landau levels belonging to different minibands is also clearly illustrated. In the zero-field semi-classical picture, all four vHS are vertical in the $V_g(B)$ map, yet this is markedly different from the experiments. Indeed, the reason for the tilting at \textbf{A} and \textbf{D} will be rationalized later. To this end, we need to first review how our semi-classical approach works.

\section*{Semi-classical approach}

\begin{figure*}[tbhp]%
\centering
\includegraphics[width=1.0\textwidth]{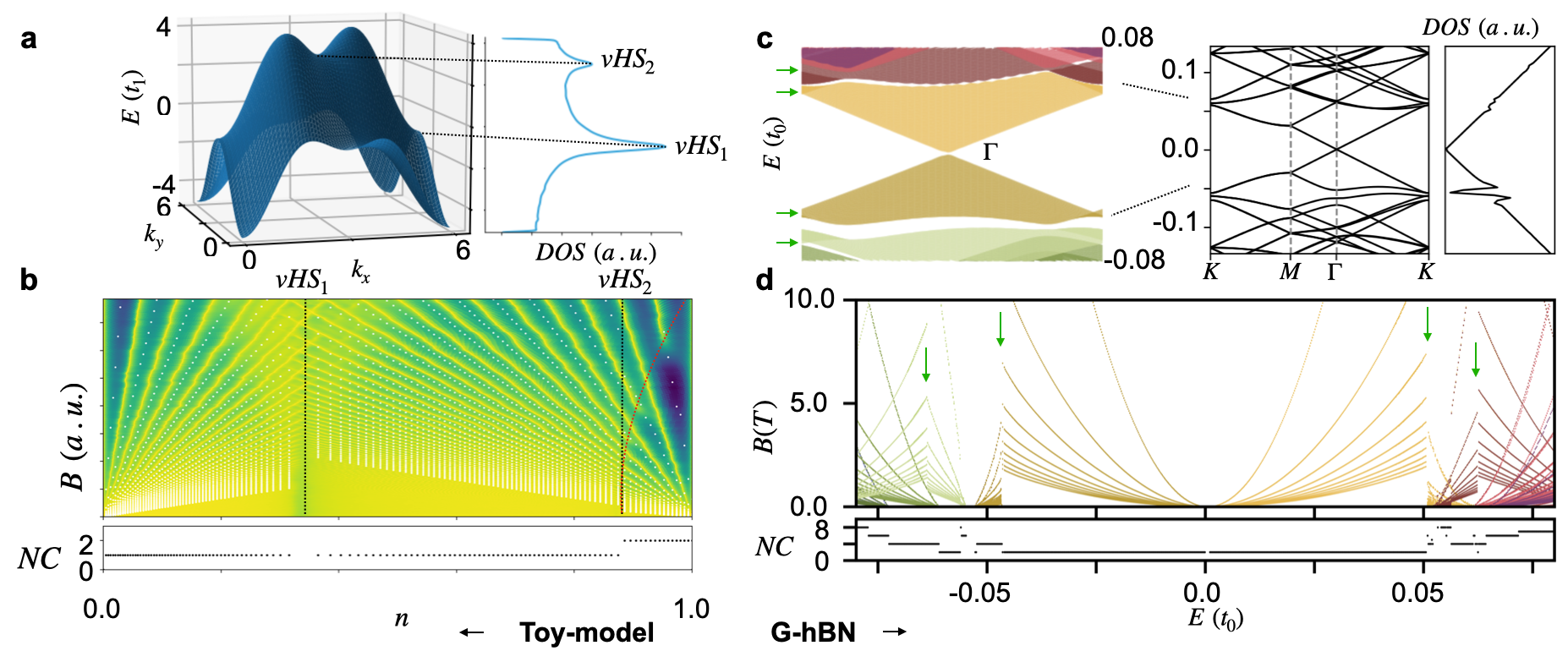}
\caption{
Semi-classical simulations highlighting Lifshitz transitions and miniband overlap. Panels \textbf{a} and \textbf{b} focus on a square-lattice single band toy model, \textbf{c} and \textbf{d} correspond to the realistic G-hBN model containing multiple minibands. \textbf{a,} Electronic band structure and its corresponding DOS. \textbf{b,} Semi-classical predictions of the LL (white dots) with the full quantum calculations in the background (dark colors correspond to a large number of states while light colors correspond to the gaps separating the LL). The dashed lines correspond to the positions of vHS where the black line based on the semi-classical model is vertical and the red line drawn based on the quantum simulations deviates to the right when the field is increasing. \textbf{c,} 3D and 2D electronic band structure and DOS, indicating the electron-hole asymmetry, and the existence of a variety of minibands potentially separated by a gap or overlapping with each other. \textbf{d,} the semi-classical LL predictions where we only plot the first 10 LLs for each energy. The different colors represent the LLs belonging to different minibands from \textbf{c}. The green arrows indicate the position of the saddle point vHS separating the electron and hole character of each miniband.
}
\label{fig1}
\end{figure*}

To properly understand the capability and limitation of our semi-classical method, pivotal for revealing the intra- and inter-miniband dynamics in the moire superlattice, we first illustrate its applicability on a single band square lattice model (see Methods for details). The model's 3D bandstructure and DOS (shown in Fig.~\ref{fig1}\textbf{a}) exhibit a vHS where the nature of this single band model changes from electrons to hole character, denoted by vHS$_1$ followed by vHS$_2$ where distinct Fermi contours coalesce into one. Employing a Marching Squares algorithm, we uniquely associate each eigenenergy with a specific Fermi contour. We then compute the area enclosed by each of these contours and semi-classically relate it to a magnetic length and magnetic field to obtain a semi-classical prediction for the LL, grounded in the bands' behavior in the absence of a magnetic field (see Methods for details). In Fig.~\ref{fig1}\textbf{b}, we finally overlay these predictions with the quantum treatment results in a charge-carrier density representation by integration of the DOS. It is important to note that the semi-classical lines predict the \textit{existence of states} rather than mirroring the \textit{absence of states} within gaps separating the Landau levels that follow the well-known $n/n_0 = t \phi/\phi_0 + s$ Diophantine relation between the normalized charge carrier density $n/n_0$ and the normalized magnetic flux $\phi/\phi_0$.
Consequently, these semi-classical predictions should be contrasted with the quantum depiction's dark-colored regions where the number of states peaks.

Overall, we observe a noteworthy congruence between the semi-classical approach premised on zero-field calculations and quantum calculations where the magnetic field is explicitly incorporated via the Peierls phase correction.  
Two regimes emerge where the concordance between semi-classical and quantum methodologies appears to diminish.
At vHS$_1$, where the band transitions from hole to electron character, the semi-classical LL predictions cannot be resolved by our algorithm, unlike with the quantum approach that shows 
intersecting and overlapping Landau level fans originating from each of the band edges. At vHS$_2$ where the white dots (dark regions) split towards lower values of $n$ into doubled features due to the Lifshitz transition 
the semi-classical perspective anticipates this occurring at a constant $n$ value (dashed dark line), irrespective of the magnetic field, while the quantum view indicates a shift towards higher $n$ values with increasing magnetic field $B$ (dashed red line). 
The latter discrepancy arises because our semi-classical approach does not directly incorporate magnetic field corrections, and hence cannot capture magnetic field-induced band renormalization.

These initial evaluations using a toy-model underscore the efficacy of the semi-classical approach, prompting us to apply the same methodology to zero-degree aligned G-hBN, as depicted on the right side of Fig.~\ref{fig1}, where, similarly, two kinds of vHS are expected to occur within each miniband. 
Firstly, each of the minibands represented in different colors in the 3D band structure in Fig.~\ref{fig1}\textbf{c} contain an electron-to-hole character transition. These occur at the saddle points indicated by the green arrows. This transition becomes more apparent from the semi-classical LL predictions in Fig~\ref{fig1}\textbf{d} (using the same single color for each miniband as in Fig~\ref{fig1}\textbf{c}) showing similar behavior as in vHS$_1$ from Fig~\ref{fig1}\textbf{b} using the toy model. Secondly, the jumps in the number of Fermi contours at each energy in the bottom panel of Fig~\ref{fig1}\textbf{d} as counted by our algorithm, suggest the occurrence of Lifshitz transitions where the topology of a miniband band changes. 
In addition to these effects that match with the single-band toy-model effects, we finally also observe in Fig~\ref{fig1}\textbf{c} and \textbf{d} several regions where minibands are overlapping (e.g. between the green-shaded bands on the hole side or between the red-shaded bands on the electron side), a requirement for the observation of inter-miniband phenomena. 
Encouraged by this diverse array of features captured by the semi-classical approach, we will delve in Figs.~\ref{fig5},~\ref{fig3} and ~\ref{fig4} into specific regions of the spectrum to explain the experimental features observed in Fig.~\ref{fig2}. We categorize them depending on the distinction if they stem from intra-miniband Lifshitz transitions or from the interaction between different minibands.

\section*{Intra-miniband Lifshitz physics}
\subsection*{Open orbits}

\begin{figure*}[tbhp]%
\centering
\includegraphics[width=1.0\textwidth]{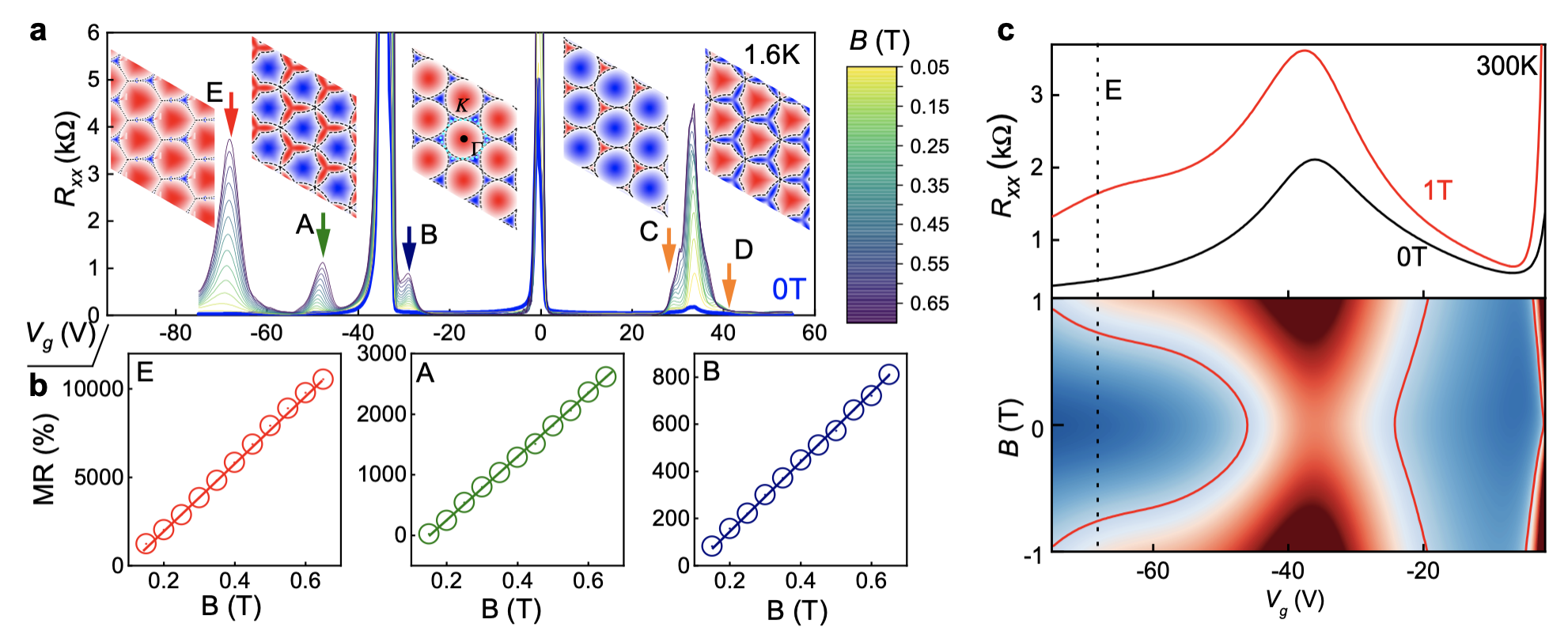}
\caption{ Open-orbit induced XMR persisting up to room temperature. \textbf{a,} $R_\text{xx}$ Rxx at low magnetic fields. The positive XMR occurs at the positions \textbf{A}, \textbf{B}, \textbf{C}, \textbf{D} and \textbf{E} where open orbits exist within each miniband. Inset: Fermi contours for each miniband showing the existence of these open orbits whenever a transition from hole to electron character of the bands occurs at a vHS in the system. We repeat the Fermi contours in a $3\times3$ tiling where we use a divergent colormap for better visualization. Red color corresponds to hole-type carriers and blue color to electron-type carriers putting the saddle point at the white colour. The dashed lines indicate as a guide to the eye the path corresponding to an open orbit. We note that for our choice in commensuration cell, the primary Dirac cones are folded back onto the $\Gamma$-point. Only one band out of 2 energy-degenerate ones is shown here. \textbf{b,} The linear fitting of the XMR of the peak resisstance values for \textbf{E}, \textbf{A} and \textbf{B}. \textbf{c,} $R_\text{XX}$ at $0$ and $1$T at room temperature where the bottom panel weakly shows the residual XMR signal at \textbf{E}. Much stronger signals are observed up to $200$K, as illustrated in the Supplemental material.}
\label{fig5}
\end{figure*}

In Fig.~\ref{fig5}\textbf{a}, we measure the XMR at low magnetic
fields from $0$T to $0.65$T where Shubnikov de Haas oscillations are not yet significant. At the saddle point vHS indicated by \textbf{A}, \textbf{B}, \textbf{C}, \textbf{D} and \textbf{E}, the resistivity increases significantly with magnetic field. The corresponding positive XMR behaviors for hole-side vHS are shown with linear fitting to magnetic fields in Figs.~\ref{fig5}\textbf{b}. Since the distances between the SDP and vHS on this hole side are larger than those the electron side (as already highlighted in Fig.~\ref{fig2}\textbf{a}), the amplitude of the peaks corresponding to the vHS are less influenced by the more significant SDF peak. The values of XMR at $1.6$K achieve $10000\%$, $3000\%$ and $800\%$ at $0.65$T for vHS \textbf{E}, \textbf{A}, and \textbf{B}, respectively. Due to \textbf{C} and \textbf{D} being closer to the SDP, their positive XMR is more prone to smearing when the temperature increases. Therefore we focus on vHS \textbf{E} when checking the temperature dependence to conclude on an XMR that can persist up to room temperature, as demonstrated in Figs.~\ref{fig5}\textbf{c}. The room temperature signal at E is very weak, but we illustrate in the Supplemental material how it strongly survives up to $200$ K.

The linearity of the XMR is a significant hallmark symbolizing the emergence of open orbits~\cite{PhysRevLett.111.056601,PhysRevB.99.035142,chambers1990electrons, pippard1989magnetoresistance, lifshits1973electron, abrikosov1988fundamentals}. Indeed, the gate voltage at which each of these occurs perfectly matches with the saddle point vHS that indicate a transition from electron to hole character~\cite{Markiewicz1994}. These transitions give rise to open orbits at zero field in reciprocal space as illustrated by the insets Figs.~\ref{fig5}\textbf{a} where the open orbits are demarcated by dashed black lines within the
white regions of the colormap. The positive XMR in open orbits can be rationalized as follows. Because a magnetic field induces a force on a moving electron, its momentum changes, and thus its path in real space also gets affected. If electrons are on closed orbits in reciprocal space, the change in moment following the k-path is periodic, and thus overall, the net change in motion in real space is negligible. Electrons move almost freely through the sample unaffected by the magnetic field. However, if they are on a closed orbit in reciprocal space, the change in momentum will be mostly in one primary direction. The magnetic field thus continually modifies the momentum of an electron at this energy in a specific direction. This leads to an increase in scattering and less predictable electron motion. The electrons don't move freely through the sample and can experience more scattering events. This in turn explains the positive XMR.
%~\footnote{visualization of currents is possible now, but parameter calibration is not solved yet. Preliminary results seem to indicate that in the presence of a magnetic field, open orbit currents follow edges of the triangles formed by moire patterns where certain edges cancel out each other's currents, while in the case of closed orbits, currents can follow many different edges quite randomly, maybe even inside the triangles.}.
By relating the observed XMR to the occurrences of open
orbits, we provide the first documentation of such features
in layered moir\'e systems. Notably, the observation of open orbits does not require
quantum mechanical treatment for validation. When comparing the energy at which
each of these open orbits occur in Figs.~\ref{fig2}\textbf{a} and \textbf{c} we nonetheless note that the zero-field
semi-classical approach does not capture their weak field-dependence due to the omittance of magnetic field-induced band renormalization. To strengthen the attribution of the XMR to open orbits, nine other similar samples with graphene doubly aligned to both sides with h-BN at various twist angles are provided in the Supplemental material showing similar observations.

\subsection*{Trigonal-warping}

\begin{figure*}[tbhp]%
\centering
\includegraphics[width=0.8\textwidth]{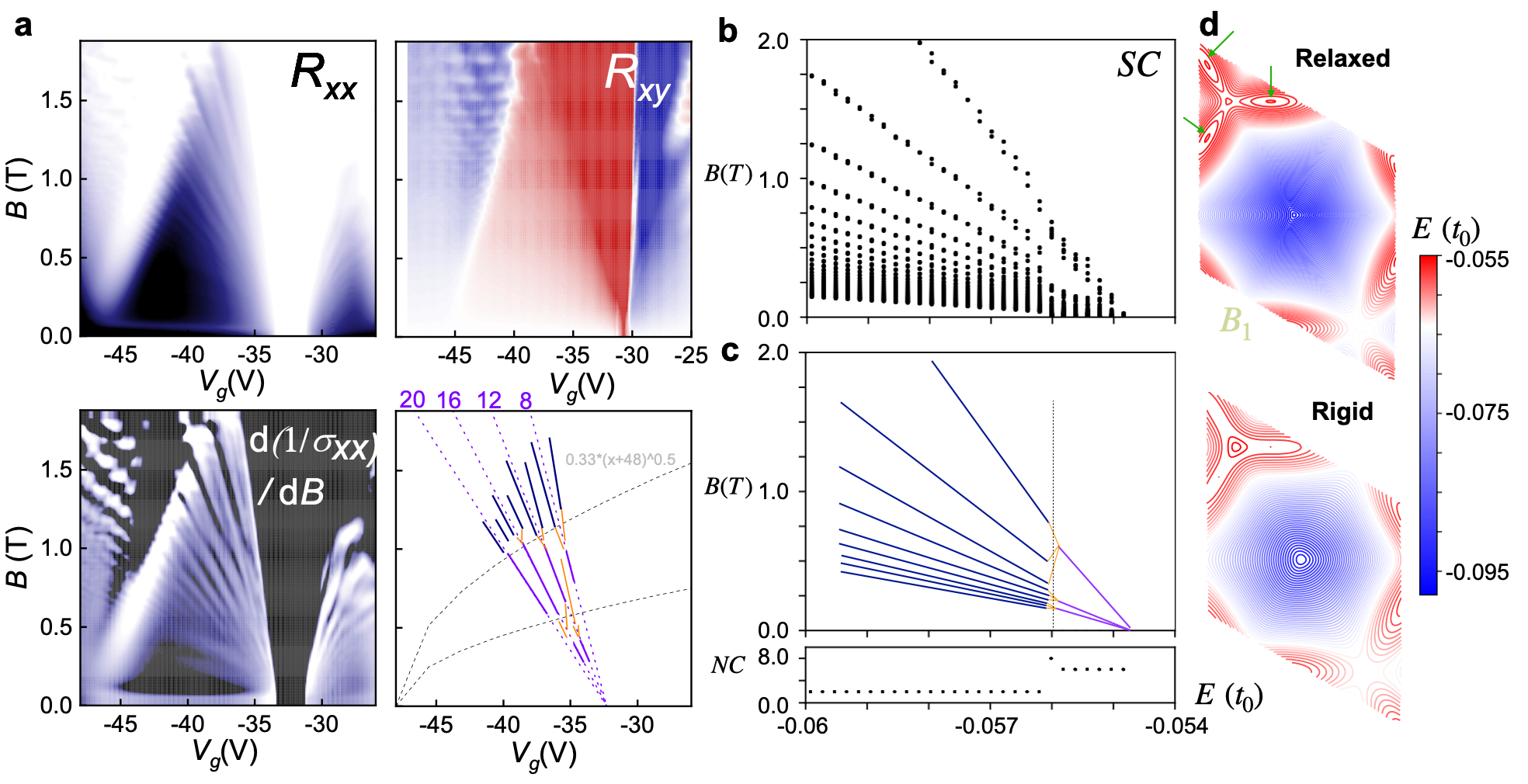}
\caption{Trigonal warping transport signatures at negative gate voltages. \textbf{a,} $R_\text{xx}$, $R_\text{xy}$, $\frac{d(1/\sigma_\text{xx})}{dB}$ measurements and the sketch based on $\frac{d(1/\sigma_\text{xx})}{dB}$ showing splitting of Landau levels. \textbf{b} Semi-classical confirmation of splitting where we plot the $30$ first LLs for each energy. \textbf{c,} sketch based on \textbf{b}. \textbf{d,} Fermi contours showing the split originates in the trigonal-warping induced merging of Fermi pockets indicated by green arrows where the relaxed model shows stronger features than when using the rigid model parameters.}
\label{fig3}
\end{figure*}

In Fig.~\ref{fig3}\textbf{a}, our attention is directed to the very low-B regime at negative gate voltage corresponding to the region demarcated by grey dashed lines in Fig.~\ref{fig2}\textbf{a} where the current ultra-high resolution measurements exhibit surprising LL level splittings at the SDP: initially, two lines split into four, eventually evolving into eight lines. This pattern can be elucidated by the semi-classical predictions in Fig.~\ref{fig3}\textbf{d} that show that at lower energy in the red-colored region of the map, three separate Fermi pockets exist around the K-point (see green arrows). These pockets merge into a single pocket as the energy slightly increases. The relaxed parameters that are used throughout this paper show clearer signatures of this effect compared to the rigid model parameters indicative of the importance of lattice reconstruction in this system. This fusion of contours that happens at a Lifshitz transtion has been associated in our toy-model with a reduction in the number of LLs where two pockets merged into one. Given that here three pockets coalesce into one, it logically follows that a single LL should split into three levels, a hypothesis supported by the semi-classical LL predictions in Fig.~\ref{fig3}\textbf{b} and the corresponding sketch in Fig.~\ref{fig3}\textbf{c}. From an experimental perspective, the progression from 2 (approximately 3) observable bands to 4 (approximately 6) and then to 8 (approximately 9) visible bands implies that, in reality, two pockets merge initially before joining with the third pocket. This sequence of events could potentially be attributed to minor non-uniform strain within the sample. The uncertainty on the exact values stems from the difficulty of extracting very fine features. As a side note, a similar observation can be made again here in terms of the semi-classical approach, namely that it ignores any magnetic field-induced renormalization of bands, missing the experimentally observed gate voltage-dependence of the splitting behavior.

The origin of these three pockets can be attributed to trigonal warping~\cite{KostyaScience}, defined as a deformation of the circular Fermi cones around the K-points that tends to be trigonal due to the lattice symmetries. Such trigonal warping usually has a substantial effect on the bandstructure of graphene, hence 
explaining its observation through ARPES measurements~\cite{PhysRevB.86.081405}. Its electronic effects on the contrary are often subtle making it challenging for transport measurements to detect them. Here by matching the observation of these subtle features in the experiments with the trigonal warping-induced deformation of this higher energy moire mini band beyond the h-SDP, we provide the first experimental confirmation of trigonal warping effects by means of transport measurements.

\section*{Inter-miniband physics}
\subsection*{Non-linear Landau levels}

\begin{figure*}[tbhp]%
\centering
\includegraphics[width=0.8\textwidth]{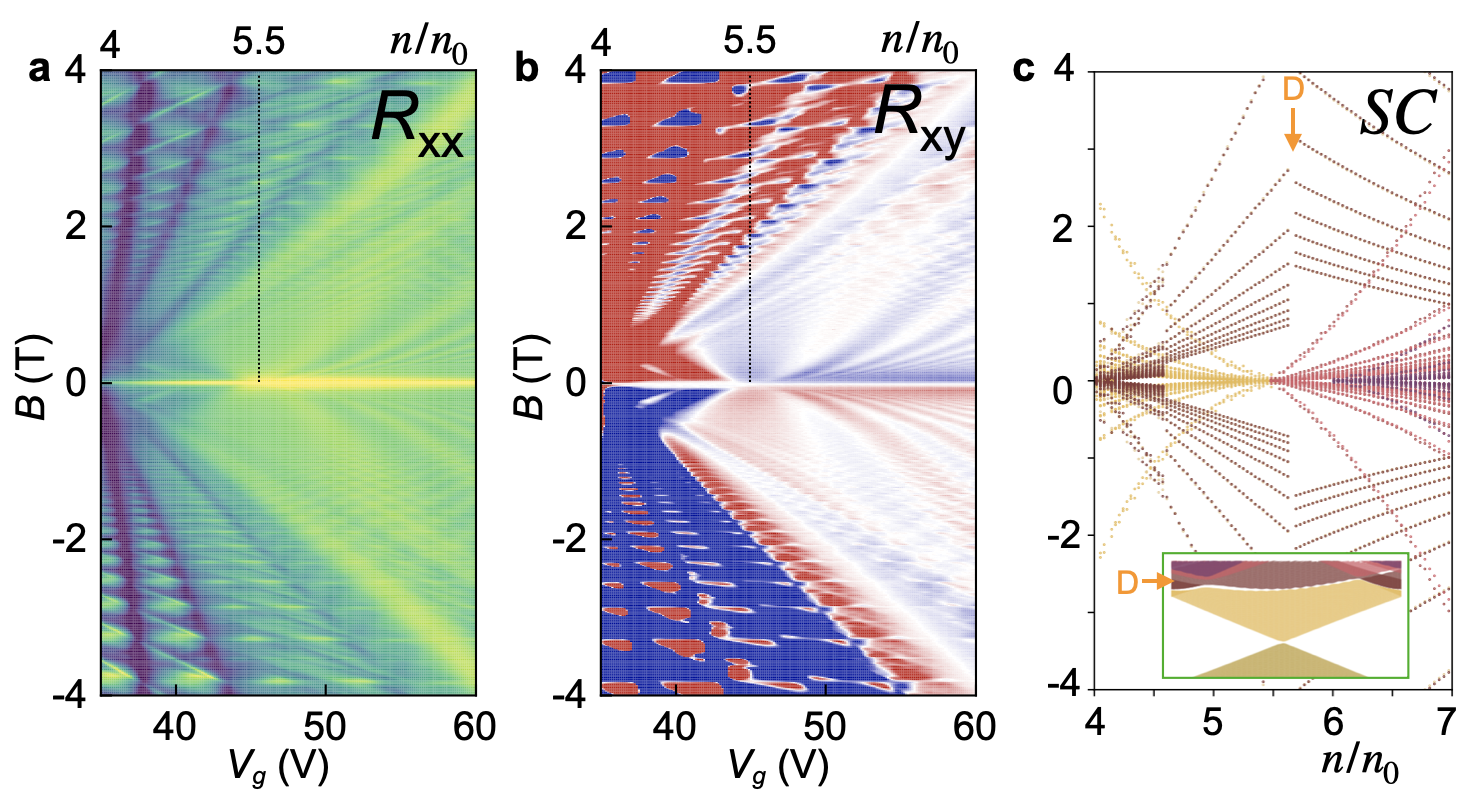}
\caption{Non-linear features and magnetic breakdown on the electron side. \textbf{a,} $R_\text{xx}$ measurements with focus on positive gate voltage beyond the e-SDP showing non-linear features emerging at $5.5 n/n_0$ both towards the left and right sides of the spectrum. \textbf{b,} Similar non-linear $R_\text{xy}$ features showing red-green color alternation indicating overlap between hole and electron character charge carriers. \textbf{c,} Semi-classical (SC) Landau level predictions confirming the non-linear behavior of the experimental features including the overlapping hole and electron character part of the bands. We only plot the ten first LLs for figure clarity. We mirror the positive magnetic field calculations to obtain the corresponding negative field bands for easier comparison with \textbf{a} and \textbf{b}. The inset shows a zoom on the relevant bands from Fig.~\ref{fig1}\textbf{c} highlighting the multitude of overlapping bands around \textbf{D}.}
\label{fig4}
\end{figure*}

Turning our attention here to inter-miniband phenomena, we focus on the measurements in Fig.~\ref{fig4} corresponding to the purple dashed regions from Fig.~\ref{fig2}\textbf{a} to investigate the departure from non-linear Landau levels in the Wannier representation of the Hofstadter butterfly. Although older works have observed such signatures, no clear origin has been provided~\cite{PhysRevB.102.045409}.% where such departure from the linear regime appears. 
The electron side features giving the clearest $R_\text{xx}$ and $R_\text{xy}$ signatures in Figs.~\ref{fig4}\textbf{a} and \textbf{b} respectively, we draw from the alternating colors corresponding to hole and electron charge carriers in $R_\text{xy}$ that miniband overlap is happening here. Fig.~\ref{fig4}\textbf{c} confirms this attribution through the dense crossing between brown, orange, pink, and purple bands in the $[4-7] n/n_0$ range
to 7 n/n0 (brown, orange, pink, and possibly purple colors in c). The colors match the colors from the minibands in the inset.

Because no magnetic field or quantum effect is explicitly included in this semi-classical approach, the non-linear effect must be explained without including quantum effects such as magnetic tunneling at the origin of magnetic breakdown. Reminding that the position of the Landau levels in the Wannier representation depends solely on the number of electrons contained within a band, the specific shape of each band does not affect the linear dependence. However, when two (mini-)bands coexist at the same energy, the filling of the two bands happens concurrently, hence while filling the first band, some electrons will start filling the second band hence disturbing the linear behavior of the first band (and vice-versa). 

\subsection*{Magnetic breakdown}
Beyond the non-linear effects in Fig.~\ref{fig4} that are already captured at zero-field, our semi-classical approach cannot capture quantum tunneling between different bands that can happen at finite magnetic field. However, the experimental maps show clear broadening of the non-linear features with magnetic field when alternating between the overlapping electron and hole character bands. This is commonly referred to as \textit{magnetic breakdown}~\cite{PhysRevLett.7.231, PhysRev.138.A88, Pippard1965, PhysRev.126.1636}. 
The quantum results for this low B-field regime are not shown, as computational constraints hinder the resolution of similar features, likely due to the requirement of exceedingly large samples for very low magnetic fields and extensive cyclotron orbits. Yet, the magnetic treatment in Fig.~\ref{fig2}\textbf{b} shows that this approach captures the experimental magnetic broadening breakdown effects in the $[1-6]$T regime hence overcoming this limitation of the semi-classical approach and confirming the quantumness of these smoothened bands. 
We finally note that experimentally, nonlinear features and magnetic breakdown are particularly pronounced in double moire systems at low-magnetic field due to hBN encapsulation of the graphene layer, as illustrated in the Supplemental material, rationalized by the fact that the additional superlattice features increase the occurrence of mini-band overlap at the root of this peculiar behavior.
As a side note, in the semi-classical picture from Fig.~\ref{fig4}\textbf{c}, the brown band originating at $4 n/n_0$ also displays a signature of trigonal warping-induced Lifshitz transitions causing the LLs to split around $4.5 n/n_0$ as was observed on the hole side in Fig.~\ref{fig3}. However, the effect is much weaker and cannot be resolved in the current measurement.

\section*{Conclusions and Outlook} 
From a fundamental point of view, our analysis of XMR measurements in graphene/hBN superlattices provides compelling evidence of fundamental limitations to accessing endless intricacies of the fractal quantum spectrum in moire superlattices, as highlighted through seminal observations of phenomena that occur in the very low B-field regime before the quantum Hall effect can fully develop. 
From an application point of view, we particularly want to highlight the open orbit-induced XMR signatures that survive up to room temperature, hence providing a promising playground to engineer similar behavior at lower gate voltages for use in magnetic field sensors, MRAM and logic gates~\cite{PhysRevB.88.195429, GANI2020126171,PhysRevB.77.081402,Friedman2010-gr,Liao2012-cg}. 
From a methodological point of view, the theoretical analysis of high-quality G-hBN samples puts in perspective the strengths and limitations inherent to semi-classical methodologies when juxtaposed with quantum mechanical calculations. The utility of semi-classical approaches, particularly evident in the elucidation of zero-field bands and Lifshitz transitions, offers a substantial contribution to understanding complex physical phenomena. In addition to the open orbit-induced XMR, these semi-classical calculations notably explain the emergence of trigonal warping-driven transport anomalies and shed light on the notable deviations from the linear B-field behavior expected for Landau levels in the Wannier representation. The full quantum picture in turn allows 
to capture magnetic breakdown due to tunneling between different bands. We expect the semi-classical approach to be useful for a variety of layered materials where Lifshitz transitions within single minibands and semi-classical and quantum interactions between different minibands can trigger a variety of exotic properties.

In the current understanding, the XMR associated with open orbits is primarily interpreted through a reciprocal space framework of electronic band structures. Our preliminary investigations, not detailed in this manuscript, indicate that in real space these open orbits correspond to localized states at the corners of the moir\'e pattern in this 2D material. This contrasts with the more spatially extended states linked to closed orbits. Introduction of a magnetic field results in a notable modulation of these spatial characteristics: states associated with open orbits exhibit a tendency towards delocalization, whereas those corresponding to closed orbits show enhanced localization due to early onset of cyclotron orbits. The observed negative XMR in the latter can still be coherently explained within the paradigm of weak localization as the localization of states is very limited at these small magnetic fields. Conversely, for open orbit states, the observed positive XMR potentially aligns with a model involving edge states. These edge states are hypothesized to interfere with each other, leading to an increase in resistivity. Future studies are planned to substantiate this hypothesis by directly visualizing, in real space, the current paths that correspond to open orbits in reciprocal space.

\section*{Methods}

\subsection*{Experiments}
\subsubsection*{Device Fabrication}
Monolayer graphene and hexagonal boron nitride with a thickness of $\sim 20$ nm were mechanically exfoliated on SiO2/Si substrates and identified under an optical microscope. The flakes with a well-defined long and straight edge were selectively utilized. 
The hBN-graphene-hBN heterostructures were assembled using standard dry-transfer technique with PMMA-coated PDMS stamps. 
To obtain the single-aligned samples, the straight edges of the top hBN and graphene were aligned to zero.
Simultaneously, the bottom hBN layer was intentionally rotated to 15 degrees to prevent the formation of unintended superlattices. By contrast, for doubly-aligned devices, all the selected straight edges of the flakes were aligned in unison. All the heterostructures were pre-examined with Raman spectrum to ensure the precise alignment. Subsequently, standard electron beam lithography was carried out to define a Hall-bar geometry using a PMMA mask, followed by CHF3/O2 ion etching. Cr/Au ($2/70$ nm) was thermally deposited to form edge contacts.
\subsubsection*{Transport measurements}

Transport measurements were performed in a dilution refrigerator with a base temperature of $30$ mK. We used standard low-frequency lock-in techniques with an excitation frequency of about $6-30$ Hz and an excitation current of about $2-5$ nA. 
The current flowing through the sample was amplified by a current pre-amplifier and measured by the lock-in amplifier. The four-probe voltage was amplified by a voltage preamplifier at $\times 1,000$ gain and measured by another lock-in amplifier. The Second Order Low Pass Filter was applied with the R1 = $512\Omega$, C1 = $4.7$ nF, R2 = $816 \Omega$, C2 = $2.2$ nF.

\subsection*{Simulations}
\subsubsection*{Semi-classical and quantum methods}
In the semi-classical method, we associate each eigenenergy with a specific Fermi contour within a specific band using a Marching Square algorithm as illustrated in the Supplemental material. We then integrate the area of these contours at zero magnetic field for each energy to obtain a semi-classical prediction of each LL for this energy.
%We summarize the main equations here and refer to the appendix for visual illustration of intermediate steps. 
The area in the case of graphene can be linked to the magnetic length $l_B$ as
\begin{equation}
    A(E,N) = 2 \pi \frac{1}{l_B(E,N)^2}\left(N+\frac{1}{2}\right)
\end{equation}
where $N$ is the Landau level number. Conversely, we have
\begin{equation}
    l_B(E,N) = \sqrt{\frac{2 \pi}{A(E)} \left(N+\frac{1}{2}\right)}.
\end{equation}
The magnetic length is then related to the magnetic field in Tesla when $l_B$ is given in nm by
\begin{equation}
B(E,N) = \left(\frac{2.6}{l_B(E,N)}\right)^2.
\end{equation}
By obtaining the magnetic field corresponding to the area for each energy and each value of $N$, we obtain the dots that are plotted in Figs.~\ref{fig1}\textbf{d}.
For the quantum calculations, the density of states (DOS) maps are obtained using Lanczos recursion or KPM~\cite{lanczos} (see Supplemental material for details) on large
systems containing on average 100 million atoms. This allows to probe very fine
meV resolution features. Quantum capacitance maps are calculated using the equations outlined in the Supplemental material. The transformation method to go from Hofstadter energy-dependent maps~\cite{PhysRevB.14.2239} to Wannier charge carrier density
n-dependent maps~\cite{Wannier1978} for both the semi-classical and quantum results is also given in the Supplemental material. An open-source Python package that extends~\cite{PhysRevResearch.2.013229} the KPM capabilities of KWANT~\cite{Groth2014} is available for download~\cite{kpmtools}.
%
%For the double-moire systems for h-BN encapsulated graphene, commensurate cells are obtained using the approach introduced in Refs.~\cite{PhysRevB.106.115410,2301.04105} based on Ref.~\cite{Hermann2012}.

%%================================%%
%% Experiments %%
%%================================%%

\subsubsection*{Tight-binding models}
%%================================%%
%% Toy-model TB %%
%%================================%%

For the toy-model calculations from Fig.~\ref{fig1}\textbf{a} and~\textbf{b}, we engineer a square lattice model that induces a change in band topology causing Lifshitz transitions as well as a transition from the hole to electron character of the charge carriers within this single band model. The first and second-neighbor interactions (see Supplemental for sketch) are given by $t_1=-1$ [$t$], $t_2=-0.5$ [$t$], $t_3=0.5$ [$t$] and $t_4=0.5$ [$t$].
%t=-1, t2=-0.5, t3=0.5
%%================================%%
%% G-BN TB %%
%%================================%%
%
%As was shown for trilayer graphene~\cite{10.1038/nphys2008} where the
%experiments were actually used to exactly determine the
%Slonczewski-Weiss-McClure parameters, 
For the G-hBN calculations, because the accuracy of the TB model is critical for our intended purpose of making qualitative comparisons with experiment, 
%in correctly reproducing the experimental Landau-level features and possible
%crossing between bands. With this in mind, 
we use a DFT-inspired TB model as introduced in Ref.~\cite{Leconte2020} based on a relaxed-model continuum parametrization~\cite{PhysRevB.91.245422}.
For the quantum treatment of the perpendicular magnetic field, the Peierls phase correction to the hopping terms is included
following the recipe introduced in Ref.~\cite{PhysRevB.103.045402}.

%\backmatter
%
%\bmhead{Data availability}
%Experimental and simulation data that support the plots within this paper are available from the first author and the last author respectively upon reasonable request. 
%
%\bmhead{Code availability}
%The codes that are used to obtain are available upon reasonable request while the semi-classical calculation methods are available as a library on our Github.
%
%\bmhead{Contributions}
%N.L., K.N. and S.R. designed the objectives of the study and led its advancement. Z.W. fabricated the devises, performed the measurements and Z.W. and K.N. analyzed the data. P.P. and N.L. performed the numerical simulations. Z.W. and N.L. wrote the manuscript with inputs from all other authors.
%
%\bmhead{Acknowledgments}

\begin{acknowledgments}
This work was supported by the Korean NRF through the Grant NRF-2021R1A6A3A13045898 (N.L.) and by the Samsung Science and Technology Foundation Grant No. SSTF-BAA1802-06 (J.J.).
We acknowledge computational support from KISTI Grant No. KSC-2022-CRE-0514 and by the resources of Urban Big data and AI Institute (UBAI) at UOS. J.J. also acknowledges support by the Korean Ministry of Land, Infrastructure and Transport (MOLIT) from the Innovative Talent Education Program for Smart Cities.
S.R acknowledges grant PCI2021-122035-2A-2 funded by MCIN/AEI/10.13039/501100011033 and European Union NextGenerationEU/PRTR, funding from the European Union's Horizon 2020 research and innovation programme under grant No 881603 (Graphene Flagship), and the support from Departament de Recerca i Universitats de la Generalitat de Catalunya. ICN2 is funded by the CERCA Programme/Generalitat de Catalunya and supported by the Severo Ochoa Centres of Excellence programme, Grant CEX2021-001214-S, funded by MCIN/AEI/10.13039.501100011033.
\end{acknowledgments}

\bibliography{all}% common bib file
%%% if required, the content of .bbl file can be included here once bbl is generated
%%%\input sn-article.bbl
%
%%% Default %%
%%%\input sn-sample-bib.tex%

\end{document}